\documentclass[preprint,prb,aps,showpacs]{revtex4}%
\usepackage[english]{babel}
\usepackage{amsfonts}
\usepackage{amsmath}
\usepackage{amssymb}
\usepackage{graphicx}%
\begin{document}
\title{Effect of Nuclear Quadrupole Interaction on the Relaxation in Amorphous Solids\\}
\author{\textit{I. Y. Polishchuk,}$^{\ast\dag}$\textit{P. Fulde,}$^{\ast}$\textit{A.
L. Burin,}$^{\ddagger}$\textit{Y. Sereda,}$^{\ddag}$\textit{D. Balamurugan}%
$^{\ddag}$}
\affiliation{$^{\ast}$Max-Planck-Institut f\"{u}r Physik komplexer Systeme, N\"{o}thnitzer
Str. 38, 01187 Dresden,Germany\linebreak$^{\dag}$Kurchatov Institute, 123182
Moscow, Russia\linebreak$^{\ddagger}$Department of Chemistry, Tulane
University,New Orleans, LA70118, USA}
\date{\today}

\begin{abstract}
Recently it has been experimentally demonstrated that certain glasses display
an unexpected magnetic field dependence of the dielectric constant. In
particular, the echo technique experiments have shown that the echo amplitude
depends on the magnetic field. The analysis of these experiments results in
the conclusion that the effect seems to be related to the nuclear degrees of
freedom of tunneling systems. The interactions of a nuclear quadrupole
electrical moment with the crystal field and of a nuclear magnetic moment with
magnetic field transform the two-level tunneling systems inherent in amorphous
dielectrics into many-level tunneling systems. The fact that these features
show up at temperatures $T<100mK$ , where the properties of amorphous
materials are governed by the long-range $R^{-3}$ interaction between
tunneling systems, suggests that this interaction is responsible for the
magnetic field dependent relaxation. We have developed a theory of many-body
relaxation in an ensemble of interacting many-level tunneling systems and show
that the relaxation rate is controlled by the magnetic field. The results
obtained correlate with the available experimental data. Our approach strongly
supports the idea that the nuclear quadrupole interaction is just the key for
understanding the unusual behavior of glasses in a magnetic field.

\end{abstract}

\maketitle

\section{Introduction}

In 1998 Strehlow et al. \cite{1} observed a pronounced but unforeseen magnetic
field dependence of the dielectric constant of certain multicomponent glasses
at ultralow temperatures $T$, i.e., below $10mK$, for fields as small as
$10\mu T.$ The effect was found to be by several orders of magnitude larger
than expected for an insulator in the absence of magnetic impurities. This
result was especially astonishing since earlier experiments did not indicate
noticeable magnetic field effects \cite{1a}. Yet, careful measurements
\cite{2} in magnetic fields ranging up to $25T$ and temperatures below $100mK$
revealed that the magnetic field causes drastic changes in the dielectric
response. These experiments have caused further investigations \cite{3,4,5}.
Following the original discovery, a number of different properties of the
glasses were investigated in magnetic fields at these temperatures
\cite{5a,5b,5c,5d,5e}, e.g., by using the dielectric dipole echo technique
\cite{dipole-echo}.

Many low-temperature properties of glasses have been successfully described in
the past by a standard tunneling model \cite{6,7,8}. To a good approximation a
Tunneling System (TS) can be treated as a particle moving in a double-well
potential (DWP). Because of the randomness of the glassy structure, the energy
difference between the two wells as well as the tunneling matrix element have
a broad distribution. This distribution is practically universal for all known
dielectric glasses and results in an agreement between theory and experiment
above $100mK$. Yet, below this temperature it is necessary to extend the model
of isolated TS's by taking into account the long-range interaction between
them in order to interpret numerous experiments \cite{10a,10,11,12,13,14}. The
concept of resonant pairs (RP) plays an important role here. It captures the
important physics as long as the interaction between TS is not so strong that
the tunneling picture looses its meaning.

After the discovery of the anomalous glass behavior in a magnetic field,
several extensions of the standard TM have been suggested \cite{8a, 8b, 9}.
The dielectric properties of glasses at low temperatures are known to be due
to the TS. It is reasonable to suppose that this is also the case for glasses
in magnetic fields. Therefore, the principal question is how the magnetic
field is acting on the TS's. In our opinion, the model of W\"{u}rger,
Fleischmann and Enss \cite{9} deserves special mentioning and attention. In
the framework of the model a direct coupling between the nuclear spin of a TS
and the magnetic field takes place, though initially it was alleged that this
possibility should be ruled out \cite{3}. However, the echo experiments
convincingly evidence the influence of the nuclear moments on tunneling
\cite{5a,5b,5c,5d}. Recently Nagel et al. \cite{5e} investigated isotope
effects in polarization echo experiments in glasses. They observed magnetic
field effects on amorphous glycerol when hydrogen, which has no electric
quadrupole moment, was substituted by deuterium, that possesses
\textit{nonzero} quadrupole moment. Thereupon, one can conclude that the
quadrupole electric moment of the TS's is the key feature responsible for the effect.

\begin{figure}
[ptb]
\begin{center}
\includegraphics[
height=1.5797in,
width=2.8578in
]%
{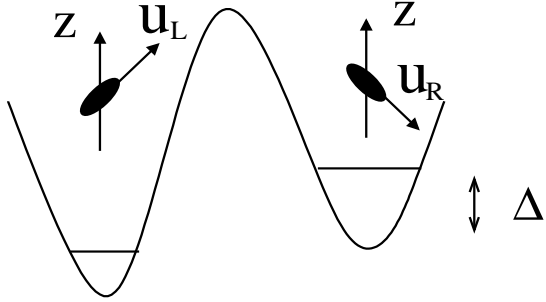}%
\caption{\textit{Tunneling system with asymmetry energy }$\Delta
,$\textit{\ and the corresponding quadrupole quantization axes in the left
well }$u_{L}$ \textit{and in the right well }$u_{R}$\textit{\ (see Ref.
\cite{9})}}%
\label{Fig1}%
\end{center}
\end{figure}

A generalization of the standard tunneling model can be done as follows
\cite{9}. Consider a tunneling particle with a nuclear spin. The energy levels
of the system are degenerate with respect to the nuclear spin projection. This
degeneracy is split if the particle has a quadrupole electric moment $Q$ which
interacts with an electric field gradient (EFG), i.e., $q_{ij}=\partial
^{2}V/\partial^{2}x_{i}x_{j},$ where $V(\vec{r})$ is the crystal-field
potential. For the sake of simplicity, W\"{u}rger et. al. \cite{9} supposed
that the EFG possesses axial symmetry. In this case, the energy of the
tunneling particle depends on the orientation of the quadrupole quantization
axis. In general, in glasses the axes $\mathbf{u}_{R}$ and $\mathbf{u}_{L}$
differ in the two wells of the potential (see Fig.\ref{Fig1}). For this
reason, the quadrupole energy changes when the particle tunnels through the
barrier. The magnetic field influences the energy spectrum since the particle
acquires a Zeeman energy depending on the nuclear spin projection.
Within that generalized tunneling model the previously mentioned echo
experiment could be described \cite{9}.

The unusual dielectric response of glasses to an applied magnetic field occurs
in the temperature region $T<50mK$ where the standard tunneling model of
non-interacting TLS cannot explain many experiments \cite{10}. Just within
this temperature region RP are responsible for the relaxation in glasses. For
this reason it is interesting to investigate whether or not the application of
the magnetic field influences the dynamics of resonant pairs.

The principal goal of the paper is to show how the quadrupole interaction, if
any, shows up in different characteristics of glasses. For this purpose we
discuss first the explanation of the anomalous magnetic-field induced behavior
of dielectric properties of glasses at ultra-low temperature based on the
model \cite{9}. Then, we clarify how these features are reflected in the
relaxation phenomena governed by the interaction between tunneling systems. We
shall extend in the paper the concept of RP so that it includes the quadrupole
interaction as well as the interaction with the magnetic field. We will also
investigate how theses interactions govern the relaxation rate of tunneling
systems. A comparison of some of the results obtained with experiments will be made.

\section{Generalization of the standard tunneling model}

Consider a double-well potential (DWP) characterized by the asymmetry energy
$\Delta.$ At low enough temperatures only two energy levels corresponding to
the ground state in each well of the DWP are significant. These levels are
connected by the tunneling amplitude $\Delta_{0}.$ According to \cite{6,7,8},
the parameters $\Delta,\Delta_{0}$ obey the universal distribution
\begin{equation}
P(\Delta,\Delta_{0})=\frac{P}{\Delta_{0}} \label{distrib}%
\end{equation}
where $P$ as a constant. An \textit{isolated} tunneling particle is usually
described by the standard two-level \textit{pseudospin} $1/2$ Hamiltonian
\begin{equation}
h=-\frac{\Delta}{2}\sigma^{z}-\frac{\Delta_{0}}{2}\sigma^{x}. \label{eq:spin}%
\end{equation}

Suppose that the tunneling particle possesses nonzero spin $\mathbf{\hat{I}%
}^{2}=I\left(  I+1\right)  $. Below we will identify it with the nuclear spin.
Then the states of the particle are characterized by the sign of the
\textit{pseudospin} projection and by the particle \textit{spin} projection
onto a proper quantization axis $n=-I,....,I$. Thus, the dimension of the
Hilbert space for the tunneling particle is $2\left(  2I+1\right)  $.
Transitions of the particle between the wells of the DWP occur with
conservation of the spin projection. Introducing the spin projection operator
$\left\vert n\right\rangle \left\langle n\right\vert $, one can define
generalized pseudospin operators $\sigma^{z}\left[  n\right]  $ and
$\sigma^{x}\left[  n\right]  $ so that $\sigma^{i}\left[  n\right]
=\sigma^{i}$ $\otimes\left\vert n\right\rangle \left\langle n\right\vert $. As
long as the spin of the particle does not interact with the environment, the
energy levels are degenerate with respect to the spin projection and the
tunneling Hamiltonian reads
\begin{equation}
h=-\sum_{n}\left(  \frac{\Delta}{2}\sigma^{z}\left[  n\right]  +\frac
{\Delta_{0}}{2}\sigma^{x}\left[  n\right]  \right)  . \label{hamnucl}%
\end{equation}
In this case, the spin degrees of freedom do not exert an influence on the
tunneling properties of the particle, no matter which Hamiltonian, either
(\ref{eq:spin}) or (\ref{hamnucl}), is used to describe the particle motion.

\begin{figure}
[ptb]
\begin{center}
\includegraphics[
height=1.7358in,
width=3.1967in
]%
{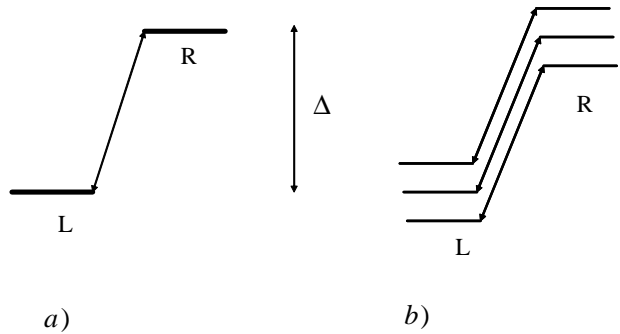}%
\caption{\textit{ Energy levels: (a) in absence and (b) in the presence of an
applied magnetic field}}%
\label{Fig2}%
\end{center}
\end{figure}

Suppose that an uniform magnetic field $B$ directed along the $z-$ axes is
applied. Then, the tunneling particle gains extra Zeeman energy dependent on
the spin projection $I_{z}$
\begin{equation}
E_{int}=g\beta BI_{z},~ \label{emag}%
\end{equation}
and the degeneracy of the energy levels is lifted. However, this splitting is
irrelevant. Indeed, in both wells of the DWP the magnetic field has the same
magnitude. For this reason, the Zeeman contribution depends only on the spin
projection and does not depend on the pseudospin projection. For the case
$I=1,$ the energy structure of the tunneling particle before and after
application of the magnetic field is presented in Fig. (\ref{Fig2}).
The states with fixed spin projection are the \emph{eigenstates}. One should
pay attention to the fact that the tunneling between the two sites $L$ and $R$
can happen only between \emph{eigenstates} that have equal spin projection
(like in the absence of a magnetic field). So, the magnetic field does not
influence the overlap integral between the wave function of the left and the
right well. This means that the application of a magnetic field alone does not
influence the properties of the TS under consideration.

Let us now consider the case when the spin of the tunneling particle is $I
\ge1$. Then the tunneling particle possesses a quadrupole electric moment. It
interacts with the crystal field which is characterized by the electric field
gradient (EFG) $q_{ij}$. The Hamiltonian of the particle interacting with the
crystal field reads \cite{15}
\begin{equation}
H_{Q}=-\frac{eQ}{2I\left(  2I-1\right)  }\left[  q_{11}I_{1}^{2}+q_{22}%
I_{2}^{2}+q_{33}I_{3}^{2}\right]  \label{f30.01}%
\end{equation}
where $H_{Q}$ is written in the basis $e_{1},e_{2},e_{3}$ in which the tensor
$q_{ij}$ has a diagonal form. The EFG satisfies Laplace's equation
$q_{11}+q_{22}+q_{33}=0.$ Introducing the asymmetry parameter
\begin{equation}
\varkappa=\frac{q_{22}-q_{33}}{q_{11}}, \label{f30.02}%
\end{equation}
one can rewrite Eq. (\ref{f30.01}) in the form
\begin{equation}
H_{Q}=-b \left(  3I_{1}^{2}+\varkappa\left(  I_{2}^{2}-I_{3}^{2}\right)
-\mathbf{I}^{2}\right)  . \label{f30.03}%
\end{equation}
Here the parameter $b=\frac{eQq_{11}}{4I(2I-1)}$ designates the quadrupole
interaction constant. We assume that the $e_{1},e_{2},e_{3}$ axes are chosen
so that $q_{33}\leq q_{22}\leq q_{11}$, since then $0\leq$ $\varkappa\leq1.$
If $\varkappa=0,$ the EFG possesses axial symmetry. In this case the
quadrupole energy is completely defined by the spin projection $I_{1}$ and the
quadrupole quantization axis is directed along $e_{1}$.

\begin{figure}
[h]
\begin{center}
\includegraphics[
height=2.2512in,
width=2.1837in
]%
{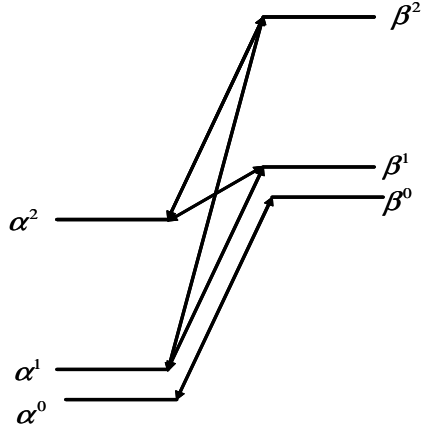}%
\caption{\textit{Energy levels structure induced by the quadrupole splitting
and possible tunneling transitions between them. In zero magnetic field the
distance between the levels with }$\left\vert \alpha^{0}\right\rangle
$\textit{ and }$\left\vert \alpha^{1}\right\rangle ,$\textit{ and the levels
\ with }$\left\vert \alpha^{0}\right\rangle $\textit{ and }$\left\vert
\beta_{1}\right\rangle $\textit{ is defined by the parameter }$\varkappa
$\textit{ and vanishes as this parameter goes to zero.}}%
\label{Fig3}%
\end{center}
\end{figure}

Assume that in the left well the basis is $e_{1},e_{2},e_{3}$ while in the
right well it is $e_{1}^{\prime},e_{2}^{\prime},e_{3}^{\prime}.$ In general,
these basises are different. Let us introduce a basis $e_{x},$ $e_{y},$
$e_{z}$ common for both wells. We assume that $e_{1}$ and $e_{1}^{\prime}$ lie
in the $e_{x},e_{y}$ plane. Suppose that $e_{1}$ coincides with $e_{x}$ while
$e_{1}^{\prime}$ forms an angle $\theta$ with the $e_{x}$ axis. Then, for the
case $I=1$ considered here, one can represent $H_{Q}$ in the \textit{right}
well in the basis of the eigenfunctions of the operator $I_{z}$ , $\left\vert
-1\right\rangle ,\left\vert 0\right\rangle ,\left\vert 1\right\rangle ,$ as
follows \cite{footnote1}
\begin{equation}
H_{Q}\left(  \theta\right)  =b\left(
\begin{array}
[c]{ccc}%
-\frac{1}{2}\left(  1+\varkappa\right)  & 0 & \frac{3}{2}\left(
1-\frac{\varkappa}{3}\right)  e^{-i\theta}\\
0 & \left(  \varkappa+1\right)  & 0\\
\frac{3}{2}\left(  1-\frac{\varkappa}{3}\right)  e^{+i\theta} & 0 & -\frac
{1}{2}\left(  1+\varkappa\right)
\end{array}
\right)  . \label{f30.0}%
\end{equation}
$H_{Q}\left(  \theta\right)  $ has the following eigenstates and eigenvalues
\begin{equation}%
\begin{array}
[c]{c}%
\left\vert \beta^{0}\right\rangle =\left\{
\begin{array}
[c]{c}%
0\\
1\\
0
\end{array}
\right\}  \leftrightarrow\varepsilon_{0}=b\left(  \varkappa+1\right)  ;\\
~\left\vert \beta^{1}\right\rangle =\left\{
\begin{array}
[c]{c}%
-e^{-i\theta}\\
0\\
1
\end{array}
\right\}  \leftrightarrow\varepsilon_{\beta^{1}}=-2b,\\
~\left\vert \beta^{2}\right\rangle =\left\{
\begin{array}
[c]{c}%
e^{-i\theta}\\
0\\
1
\end{array}
\right\}  \longleftrightarrow\varepsilon_{\beta^{2}}=b\left(  1-\varkappa
\right)
\end{array}
\label{eigenq}%
\end{equation}
For the \textit{left} well the eigenvalues are the same as for the right well
while the eigenvectors are $\left\vert \alpha^{0}\right\rangle ,\left\vert
\alpha^{1}\right\rangle ,\left\vert \alpha^{2}\right\rangle $, resulting from
Eq.(\ref{eigenq}) for $\theta=0$. It follows directly from (\ref{eigenq}) that
the spin state $\left\vert \alpha^{0}\right\rangle =\left\vert \beta
^{0}\right\rangle =\left\vert 0\right\rangle $ is orthogonal to all the
others, but because of the fact that the quadrupole quantization axes differ
in the two wells, $\left\langle \alpha^{i}\right\vert \left.  \beta
^{j}\right\rangle \neq0,~\ i\neq j$. Thus, the state with spinor $\left\vert
0\right\rangle $ in the left well can couple only to the state with spinor
$\left\vert 0\right\rangle $ in the right well, while the remaining states in
the left and the right well couple to each other. The energy level structure
and possible transitions between the left and the right well states are shown
in Fig. (\ref{Fig3}).

\begin{figure}
[ptbh]
\begin{center}
\includegraphics[
height=2.1846in,
width=2.9004in
]%
{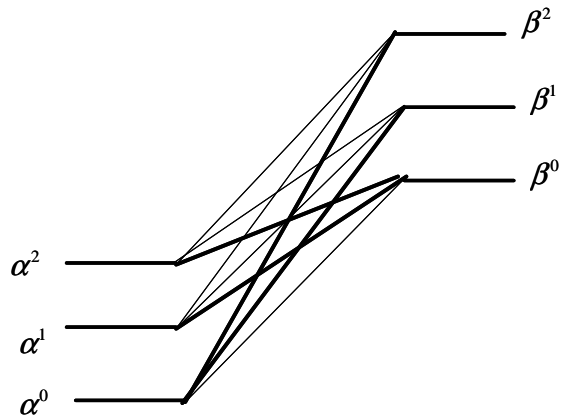}%
\caption{\textit{Shown in the figure by thick lines are the\ extra transitions
created by the application of the magnetic field. The thin lines correspond to
the previous transitions (see Fig. (\ref{Fig3})) .}}%
\label{Fig4}%
\end{center}
\end{figure}

Let us investigate how the application of the external magnetic field changes
the energy spectrum of the tunneling particle. The magnetic field shows up in
two ways. First, it produces a Zeeman splitting. Then for different TS's the
planes formed by the vectors $e_{1},e_{1}^{\prime}$ are randomly oriented. For
this reason the magnetic field is in general not orthogonal to this plane.
Therefore it mixes the states with spinor $\left\vert 0\right\rangle $ with
states with spinors $\left\vert \alpha^{i}\right\rangle $ and $\left\vert
\beta^{i}\right\rangle $ ($i=1,2$) of the two wells. In a new basis
$\left\vert \alpha^{i}\right\rangle $ and $\left\vert \beta^{i}\right\rangle $
which is dependent on the direction of the magnetic field \textit{all}
$\left\langle \alpha^{i}\right\vert \left.  \beta^{j}\right\rangle
\neq0,\ \ i,j=0,1,2$. Thus, additional transitions between the two wells
become possible (see Fig. (\ref{Fig4})).

It should be stressed that with increasing magnetic field, the Zeeman
splitting will eventually exceed the quadrupolar one. Therefore, we discuss
first the case when the TS's are affected only by the magnetic field (see Fig.
(\ref{Fig2})).

To simplify the further analysis, suppose that the magnetic field is
orthogonal to the plane $e_{1},e_{1}^{\prime}$ and that the quadrupole
splitting constant $b$ is the same in both wells. Then the Hamiltonian of the
TS in the presence of the quadrupole and Zeeman splitting reads
\begin{equation}
\left(
\begin{array}
[c]{cc}%
H_{L} & \Delta_{0}\cdot\mathbf{I}\\
\Delta_{0}\cdot\mathbf{I} & H_{R}%
\end{array}
\right)  , \label{master}%
\end{equation}
where\textbf{ }$\mathbf{I}$ is the rank $3$ unit matrix and
\begin{align}
H_{L}  &  =\left(
\begin{array}
[c]{ccc}%
-\frac{\Delta}{2}-m-\frac{b}{2}\left(  1+\varkappa\right)  & 0 & \frac{3}%
{2}\left(  1-\frac{\varkappa}{3}\right) \\
\frac{3}{2}\left(  1-\frac{\varkappa}{3}\right)  & -\frac{\Delta}{2}+b\left(
\varkappa+1\right)  & 0\\
0 & 0 & -\frac{\Delta}{2}+m-\frac{b}{2}\left(  1+\varkappa\right)
\end{array}
\right) \label{master1}\\
& \nonumber\\
H_{R}  &  =\left(
\begin{array}
[c]{ccc}%
\frac{\Delta}{2}-m-\frac{b}{2}\left(  1+\varkappa\right)  & 0 & \frac{3}%
{2}\left(  1-\frac{\varkappa}{3}\right)  e^{-i\theta}\\
0 & \frac{\Delta}{2}+b\left(  \varkappa+1\right)  & 0\\
\frac{3}{2}\left(  1-\frac{\varkappa}{3}\right)  e^{i\theta} & 0 &
\frac{\Delta}{2}+m-\frac{b}{2}\left(  1+\varkappa\right)
\end{array}
\right)  ~~. \label{master2}%
\end{align}
\newline

\section{Permittivity of a tunneling system with quadrupole and Zeeman
splitting}

Let us investigate how the changes of the energy spectrum induced by the
quadrupole and Zeeman splitting described by Hamiltonian (\ref{master})
influence the properties of a TS. First we examine the influence of these
changes on the dielectric permittivity.

Consider a particle that can occupy $(2I+1)$ levels in a DWP. We introduce the
dipole moment operator in the form
\begin{equation}
\hat{p}=\frac{p_{0}}{2}\left(
\begin{array}
[c]{cc}%
-\mathbf{I} & 0\\
0 & \mathbf{I}%
\end{array}
\right)  \label{a1.2}%
\end{equation}
where $\mathbf{I}$ is the unit matrix of rank $(2I+1)$ and $p_{0}$ defines the
value of the dipole moment. (For $I=0$ this expression transforms into the
well known dipole moment operator for a two-level-system). The interaction of
the TS with the external electrical field $F$ reads
\begin{equation}
\hat{V}=-F\hat{p}~~. \label{a1.3}%
\end{equation}
In second-order perturbation theory the correction to the energy of a TS
induced by this interaction is given by
\begin{equation}
\delta\varepsilon=F^{2}Z^{-1}\sum\limits_{b\neq a}\frac{e^{-\frac{E_{a}}{T}%
}\left\vert \left\langle a\right\vert \hat{p}\left\vert b\right\rangle
\right\vert ^{2}}{E_{a}-E_{b}} \label{a1.5}%
\end{equation}
where $T$ is the temperature, $Z=\sum\limits_{a}e^{-\frac{E_{a}}{T}}$ is the
partition function and $E_{a}$ denotes the eigenvalues of the TS in the
presence of a magnetic field.

On the other hand, as the electric field changes from zero to a certain final
value $F,$ the energy increases by $\chi F^{2}/2,$ where $\chi$ is a
permittivity of the system. Comparing this value to $\delta\varepsilon$ one
obtains
\begin{equation}
\chi=2Z^{-1}\sum\limits_{b\neq a}\frac{e^{-\frac{E_{a}}{T}}\left\vert
\left\langle a\right\vert \hat{p}\left\vert b\right\rangle \right\vert ^{2}
}{E_{b}-E_{a}}=Z^{-1}\sum\limits_{b\neq a}\frac{\left(  e^{-\frac{E_{a}}{T}%
}-e^{-\frac{E_{b}}{T}}\right)  \left\vert \left\langle a\right\vert \hat
{p}\left\vert b\right\rangle \right\vert ^{2}}{E_{b}-E_{a}} \label{a1.6}%
\end{equation}
Let us apply this relation for the TS described by Hamiltonian (\ref{master}).

The permittivity of a tunneling system is a function of the parameters
$\Delta,\Delta_{0},b,\theta,m.$ (For a preliminary analysis in this part of
the paper we confine ourselves to the case of an axial symmetric EFG by
setting the parameter $\varkappa=0$). Keeping in mind a possible fit of the
available experimental data, we choose the following values for the parameters
$\Delta_{\max}=\Delta_{0\max}=10K,$ $b=10mK.$ The temperature varies between
$10mK$ and $100mK.$ The Zeeman energy of the nuclear spin ranges from zero to
$30mK.$ With the Land\'{e} factor $g=3,$ this corresponds to a maximal
magnetic field of about $15T.$

The permittivity has been estimated numerically. First, we calculated the
eigenvalues and eigenfunctions for the matrix (\ref{master}) with values of
the TS parameters in the above mentioned range. Then, we estimated the
permittivity $\chi\left(  \Delta,\Delta_{0},b,\theta,m,T\right)  $ by using
Eq.(\ref{a1.6}). Finally, the result was averaged over the parameters
$\Delta,\Delta_{0},\theta$ by using Eq.(\ref{distrib}) and assuming a uniform
distribution of the angles formed by the quadrupole quantization axes in the
wells. As an example, we present here the result of our calculations for the
temperature $T=40 mk$%

\begin{figure}
[ptbh]
\begin{center}
\includegraphics[
height=2.4401in,
width=3.1275in
]%
{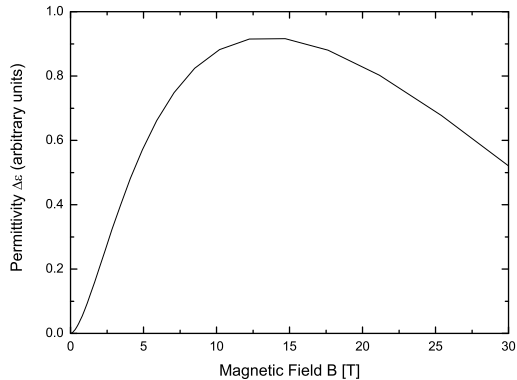}%
\caption{\textit{Magnetic field dependence of the permittivity }%
$\varepsilon^{\prime}$\textit{\ for typical quadrupole splitting }%
$b=10mK$\textit{\ at }$T=40mK$}%
\label{Fig05}%
\end{center}
\end{figure}

The permittivity exhibits a pronounced peak at an energy approximately
corresponding to the magnetic field $B\approx15T.$ This result corresponds
well to Ref. \cite{2} where a pronounced peak in the permittivity was observed
for a similar value of the magnetic field and a temperature of $T=64 mK.$

\begin{figure}
[h]
\begin{center}
\includegraphics[
height=2.6547in,
width=3.4007in
]%
{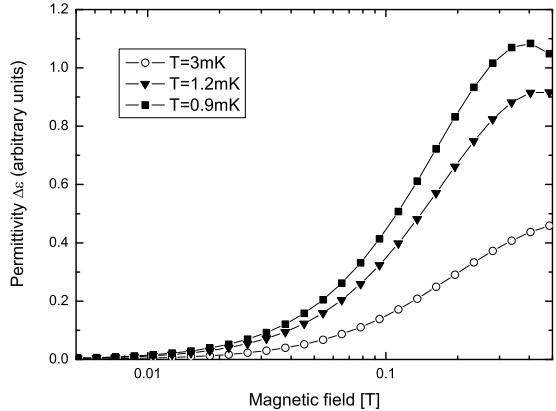}%
\caption{\textit{Magnetic field variation }$\Delta\varepsilon$\textit{ of the
permittivity.}}%
\label{fig006}%
\end{center}
\end{figure}

As mentioned before we have assumed $\varkappa=0$ in calculating the
permittivity shown in Fig. \ref{Fig05}. When we introduce $\varkappa$ a second
energy scale appears when $\varkappa$ is small. This scale is relevant for
weak magnetic fields. It might explain another peak observed in Ref. \cite{2}
in the low field regime. Also a second atom of the tunneling entity with a
different quadrupole moment may result in a second energy scale. We have
modelled a low energy scale by simply redoing the calculations for $b=0.3mK$
and plotting the results for the permittivity in a form which can directly
compared with the experimental findings of Ref. \cite{2}. Figure \ref{fig006}
shows an agreement with experiments that the permittivity at fixed value of
magnetic field is the higher the lower the temperature is.

\section{Relation between the permittivity and many-body relaxation}

As mentioned in the introduction, the temperature region $T<100 mK$ where
certain glasses show an unusual response to an applied magnetic field
coincides with the region where the relaxation of tunneling system is due to
the $R^{-3}$ interaction between TLS rather then single phonon processes. In
the previous section we have seen that a magnetic field affects the
permittivity provided the quadrupole effect is taken into account. Therefore,
it is of interest to investigate whether a magnetic field influences the
relaxation induced by this interaction.

The relaxation induced by the long range $R^{-3}$ interaction is strongly
connected with the concentration of resonant pairs (RP). Resonant pairs of
TS's are the main concept for this relaxation mechanism. First we recall
briefly the main idea of this approach.

\begin{figure}
[h]
\begin{center}
\includegraphics[
height=2.8862in,
width=3.3555in
]%
{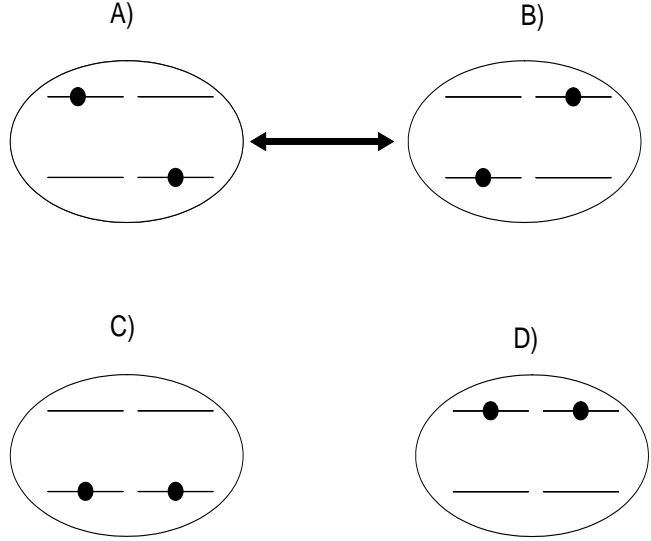}%
\caption{\textit{The states A) and B) belong to the flip - flop configuration
and form a resonant pair. The states C) and D) are energy-wise forbidden}}%
\label{Fig8}%
\end{center}
\end{figure}

Consider a pair of two-level-systems. In general, it possesses four different
configurations. Only two of them (configurations $A$ and $B$ ) shown in Fig.
(\ref{Fig8}) are important. The special feature of these configuration is that
one TLS\ is in the ground state while the other is in the excited state.

Let $V_{12}=U/R^{3}$ denote the weak interaction between these TLS's where $U$
is the interaction constant. Such a pair is in resonance when
\begin{equation}
\mid E_{1}-E_{2}\mid<V_{12} ~~. \label{f5}%
\end{equation}
Here $E_{i}=\sqrt{\Delta_{0i}^{2}+\Delta_{i}^{2}},$ $\Delta_{0i},\Delta
_{i},i=1,2$ are the tunneling parameters of two TLS constituting a pairs. We
are interested in the case $E_{1}\approx E_{2}\approx T$ and $V_{12}\ll T$
which means a weak interaction. Due to the constraint (\ref{f5}), the states
$(A)$ and $(B)$ are coupled with each other and are well separated from the
states $(C)$ and $(D)$ by an energy gap of the order of $T$. For this reason,
they can be excluded from the consideration. The two states of the
\textit{TLS} pair $(A)$ and $(B)$ are referred to as a flip-flop configuration
and the transition between them is called a flip-flop one. When such two
\textit{TLS's} form a resonant pair (RP) they can be considered as a new type
of the two - level system with the asymmetry $\mid E_{1}-E_{2}\mid$. Resonant
pairs are responsible for the many-body relaxation induced by the $R^{-3}$
interaction. The transition amplitude between the levels $(A)$ and $(B)$ is
described by the following expression \cite{10a, 11}
\begin{equation}
U_{12}=V_{12}\sigma_{1}^{x}\sigma_{2}^{x}\frac{\Delta_{01}\Delta_{02}}
{E_{1}E_{2}}. \label{trtls}%
\end{equation}
Next we write down the expression for the concentration of pairs with
tunneling parameters $\Delta_{p},\Delta_{0p}$ \cite{10a, 11}:
\begin{equation}%
\begin{array}
[c]{c}%
P^{\left(  2\right)  }\left(  \Delta_{p},\Delta_{0p}\right)  =P^{2}\int
\frac{d\Delta_{01}d\Delta_{1}}{\Delta_{01}}\cdot\int\frac{d\Delta_{02}
d\Delta_{2}}{\Delta_{02}}\left[  1+e^{-\frac{E_{1}}{T}}\right]  ^{-1}\left[
1+e^{\frac{E_{2}}{T}}\right]  ^{-1}\cdot\\
\int d^{3}R\delta\left(  \Delta_{0p}-\frac{\Delta_{01}\Delta_{02}}{E_{1}E_{2}
}\frac{U}{R^{3}}\right)  \delta\left(  \Delta_{p}-\left(  E_{1}-E_{2}\right)
\right)
\end{array}
\label{3}%
\end{equation}
The integration over $d^{3}R$ gives
\begin{equation}
\int d^{3}R\delta\left(  \Delta_{0p}-\frac{\Delta_{01}\Delta_{02}}{E_{1}E_{2}
}\frac{U}{R^{3}}\right)  =\frac{\Delta_{01}\Delta_{02}}{E_{1}E_{2}}\frac
{U}{\Delta_{0p}^{2}}. \label{4}%
\end{equation}
For resonant pairs one has $\Delta_{p}\leq\Delta_{0p}\ll E_{1}\approx
E_{2}\approx T$ and, therefore, one can omit $\Delta_{p}$ in the argument of
the $\delta-$ function in (\ref{3}). The concentration of resonant pairs
$P_{r}^{\left(  2\right)  }\left(  \Delta_{0p}\right)  $ is obtained by
integration of $P^{\left(  2\right)  }\left(  \Delta_{p},\Delta_{0p}\right)  $
over the interval $0<\Delta_{p}<\Delta_{0p}$
\begin{equation}
P_{r}^{\left(  2\right)  }\left(  \Delta_{0p}\right)  =P^{2}\frac{U_{0}
}{\Delta_{0p}}\int\frac{d\Delta_{01}d\Delta_{1}}{\Delta_{01}}\cdot\int
\frac{d\Delta_{02}d\Delta_{2}}{\Delta_{02}}\frac{1}{ch^{2}\frac{E_{1}}{2T}
}\left(  \frac{\Delta_{01}\Delta_{02}}{E_{1}E_{2}}\right)  \delta\left(
E_{1}-E_{2}\right)  \label{6}%
\end{equation}

\begin{figure}
[ptb]
\begin{center}
\includegraphics[
height=3.5674in,
width=4.6079in
]%
{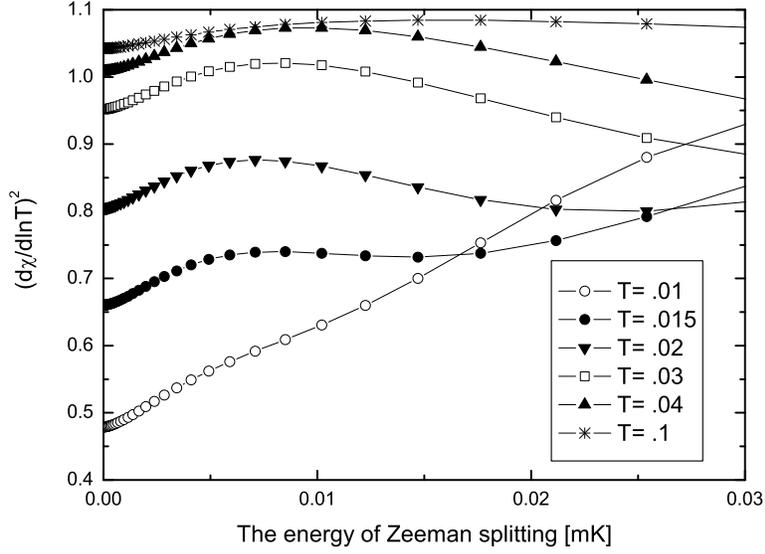}%
\caption{The square of the logarithmic derivative of the permittivity as a
function of the magnetic field.}%
\label{fig07}%
\end{center}
\end{figure}

On the other hand, the resonance permittivity of a TLS is given by the well
known expression (see e.g. \cite{7})
\begin{equation}
\chi=P\int_{0}^{B}d\Delta\int_{\delta_{0}}^{A}\frac{d\Delta_{0}}{\Delta_{0}%
}\frac{1}{E}\left(  \frac{\Delta_{0}}{E}\right)  ^{2}\tanh\frac{E}{2T}
\label{7}%
\end{equation}
Let us calculate the derivative
\begin{equation}
\frac{\partial\chi}{\partial\ln T}=-\frac{P}{2T}\int_{0}^{B}d\Delta
\int_{\delta_{0}}^{A}\frac{d\Delta_{0}}{\Delta_{0}}\frac{1}{ch^{2}\frac{E}%
{2T}}\left(  \frac{\Delta_{0}}{E}\right)  ^{2} \label{8}%
\end{equation}
Comparing expressions (\ref{6}) and (\ref{8}), we can conclude that
$P_{r}^{\left(  2\right)  }\left(  \Delta_{0p}\right)  $ and the square of
$\frac{\partial\chi}{\partial\ln T}$ are approximately proportional to each
other. Therefore we expect similar features in the behavior of the
permittivity and in the rate of the interaction induced relaxation which is
proportional to the RP concentration. This means that the multilevel systems
can again be considered as effective two level systems. When the Zeeman
splitting exceeds the quadrupole one, the multilevel tunneling systems behaves
exactly as two-level-systems. In Figure \ref{fig07} we present in the
framework of the model described above the results of the calculation of the
parameter $\left(  \frac{\partial\chi}{\partial\ln T}\right)  ^{2}$ for
different temperatures as a function of the applied magnetic field.
One notices that for magnetic fields causing a Zeeman splitting smaller or of
the order of the quadrupole splitting, i.e., $b\lesssim10mK$ it holds that the
larger the temperature the larger is $\partial\chi/\partial\ln T$. This can be
understood as follows: The relaxation rate induced by the interaction between
TS's is proportional to the concentration of RP, which in turn is proportional
to the temperature. This conclusion agrees with the sequence of different
curves in Fig.~\ref{fig07}. However, some curves show a monotonous magnetic
field dependence, while others display a non-monotone behavior. This feature
is explained in the following section.

\section{Relaxation of the Interacting many-level tunneling systems}

The main goal of the current section is a qualitative treatment of the results
obtained in Sec. IV. In particular, we want to explain the temperature
dependencies shown by Fig.~\ref{fig07}.

In our previous papers \cite{10a, 11, 12, 13, 14} we have demonstrated that
the many-body $R^{-3}$ interaction between TLS's results in a new relaxation
mechanism responsible for the low temperature relaxation. On the other hand,
it was shown above that if the tunneling particle possesses a nuclear spin
$I\geq1,$ the energy spectrum of a tunneling system consists of several lines,
as distinct from one line in the case of a TLS. Below we investigate how these
changes in the energy spectrum of the tunneling system caused by the
interaction of the tunneling particle with the electric field gradient and the
magnetic field influence the relaxation rate produced by the $R^{-3}$ interaction.

The model under investigation assumes that the particle can occupy $n=2I+1$
levels each in the left and in the right well of a TS. Let us generalize the
concept of resonant pairs to the case $n>1.$
\begin{figure}
[ptbh]
\begin{center}
\includegraphics[
trim=0.000000in -0.020211in 0.000000in 0.020211in,
height=3.0911in,
width=4.112in
]%
{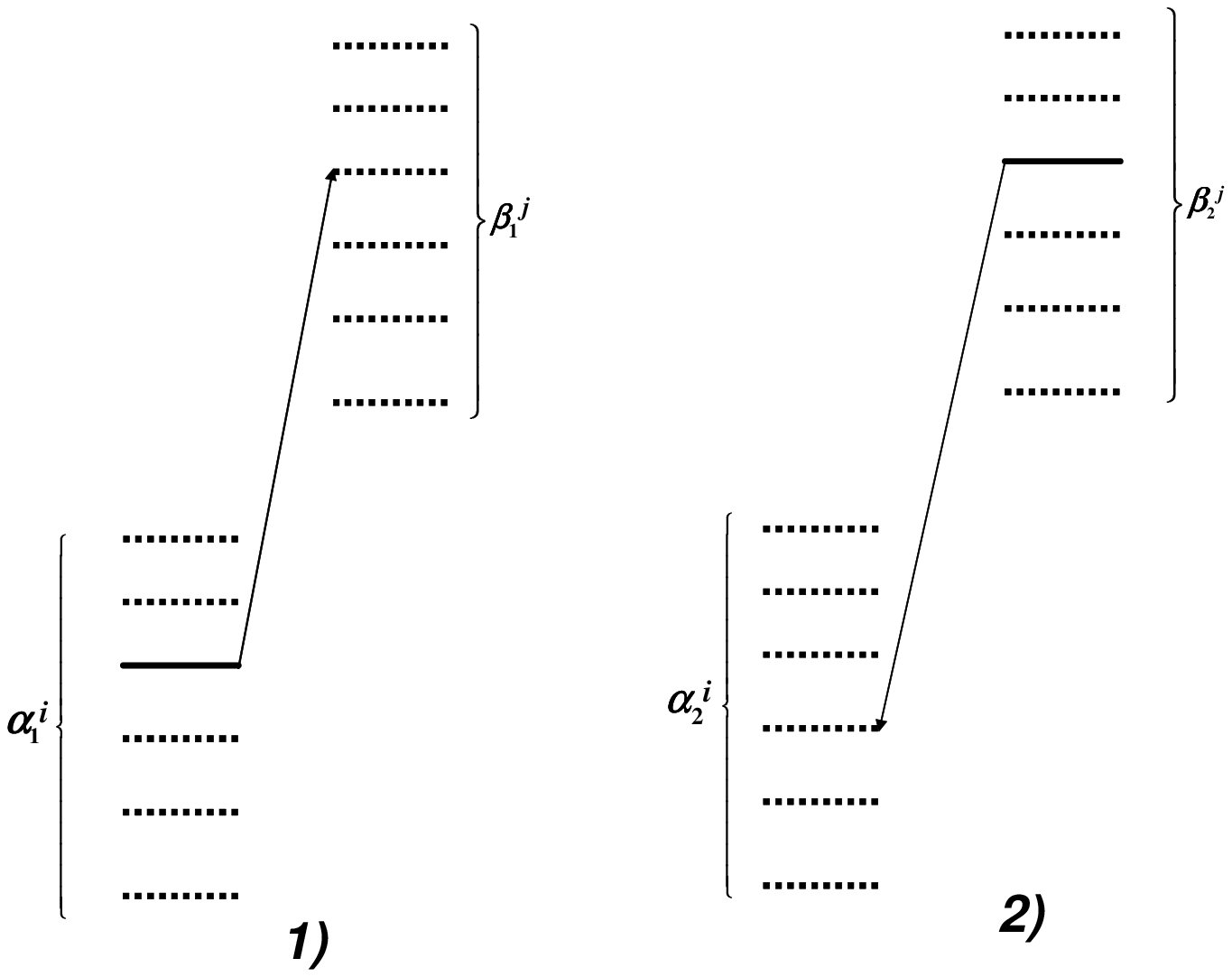}%
\caption{\textit{Two tunneling systems:TS 1) is in one of its ground state
multiplet. TS 2) is in the excited state multiplet. \newline The occupied
states are marked by solid line. The arrow lines show which levels are
occupied after flip-flop process has happened}}%
\label{fig9}%
\end{center}
\end{figure}
Consider the tunneling system (1) in Fig. \ref{fig9}. There we are dealing
with a ground-state multiplet the states of which are denoted by their spinor
parts $\alpha_{1}^{i}$ and an excited state multiplet denoted by $\beta
_{1}^{j}$. The corresponding energies are $\delta_{\alpha_{1}^{i}}$ and
$\delta_{\beta_{1}^{i}}$. They are derived from the energies $0$ and
$E_{1}=\sqrt{\Delta_{10}^{2}+\Delta^{2}}$ of the TS in the absence of the
quadrupolar and Zeeman interaction. Therefore the various transition energies
between the two multiplets are
\begin{equation}
E_{1}^{ij}=E_{1}+\left(  \delta_{\alpha_{1}^{i}}-\delta_{\beta_{1}^{j}%
}\right)  ~~,~~i,j=1,...,n~~. \label{9}%
\end{equation}
Like in the case of a TLS only those with $E_{1}^{ij}\approx T$ contribute
significantly to relaxation processes. Below we will assume that also
\begin{equation}
\delta_{\alpha_{1}^{i}},\delta_{\beta_{1}^{j}}\ll T. \label{small}%
\end{equation}
holds \cite{opposite}. Therefore only $E_{1}$ defines the Gibbs distribution
and the states within a multiplet are occupied with nearly equal probability.
A similar analysis holds for the TS(2) of Fig. \ref{fig9}. Thus, two
multilevel TS's give raise to flip-flop processes $\alpha_{1}^{i},\beta
_{2}^{j}\Rightarrow\beta_{1}^{k},\alpha_{2}^{l}$ when the corresponding energy
differences between the two configurations differ in energy by less than a
characteristic interaction matrix element. The latter and the various
different processes are discussed in the following.

Let $V\left[  \alpha^{i}_{1}, \beta^{j}_{2}; \beta^{k}_{1}, \alpha^{l}_{2}
\right]  $ denote the transition amplitude of such a flip-flop transition.
Then the condition for a resonant pair is
\begin{equation}
\left\vert E_{1}^{ij} - E_{2}^{kl} \right\vert = \left\vert E_{1} - E_{2} +
\left(  \delta_{\alpha_{1}^{i}} -\delta_{\beta_{1}^{j}} \right)  - \left(
\delta_{\alpha_{2}^{l}} - \delta_{\beta_{2}^{k}} \right)  \right\vert < V
\left[  \alpha_{1}^{i}, \beta_{2}^{j}; \beta_{1}^{k}, \alpha_{2}^{l} \right]
~. \label{newrc}%
\end{equation}

When all the energy splittings $\delta=0$, i.e., for TLS's the resonating pair
concentration is given by \cite{10a,11,12,13,14}
\begin{equation}
N_{\ast}=\left(  PT\right)  \left(  PU\right)  ~~. \label{10}%
\end{equation}

Let us investigate how the concentration of resonating pairs is modified for
multilevel systems. For simplicity we set all $V\left[  \alpha_{1}^{i}%
,\beta_{2}^{j};\beta_{1}^{j},\alpha_{2}^{i}\right]  \cong V$. We fix the
energy $E_{1}$ and also the splitting energies $\delta_{\nu}^{m}$. Then the
energy range of the parameter $E_{2}$ is defined by Eq. (\ref{newrc}) and
equals $V$. For a given $E_{1}$ there are $n^{4}$ different state
configurations $\alpha_{1}^{i},\beta_{1}^{j};\alpha_{2}^{i},\beta_{2}^{j}$
possible, i.e., the energy difference $\left(  \delta_{\alpha_{1}^{i}}%
-\delta_{\beta_{1}^{j}}\right)  -\left(  \delta_{\alpha_{2}^{i}}-\delta
_{\beta_{2}^{j}}\right)  $ takes $n^{4}$ different values. So there are also
$n^{4}$ different energy ranges for $E_{2}$. Different values of $E_{2}$
relate to different tunneling systems. However, one must take into account
that the probability of finding an initial configuration $\alpha_{1}^{i}%
,\beta_{1}^{j}$ is $1/n^{2}$. Hence the total probability of forming
resonating pairs would seem to increase like $n^{4}/n^{2}=n^{2}$ as compared
with TLS's. This would hold true if the transition amplitude $V\left[
\alpha_{1}^{i},\beta_{2}^{j};\beta_{1}^{j},\alpha_{2}^{i}\right]  $ would
coincide with the previously discussed amplitude $V_{12}$ of TLS's. But this
is not so as is easily seen as follows. The transition between two TS's $(1)$
and $(2)$ can be described similarly as in Eq. (\ref{trtls}) by
\begin{equation}
\sum\limits_{\alpha_{1}^{i},\beta_{2}^{j};\alpha_{2}^{i},\beta_{1}^{j}%
}V\left[  \alpha_{1}^{i},\beta_{2}^{j};\beta_{1}^{j},\alpha_{2}^{i}\right]
\sigma_{1}^{x}\sigma_{2}^{x}\left\vert \alpha_{1}^{i}\right\rangle \left\vert
\beta_{2}^{j}\right\rangle \left\langle \alpha_{2}^{i}\right\vert \left\langle
\beta_{1}^{j}\right\vert ,\label{mh}%
\end{equation}%
\begin{equation}
V\left[  \alpha_{1}^{i},\beta_{2}^{j};\beta_{1}^{j},\alpha_{2}^{i}\right]
=V_{12}\left\langle \beta_{1}^{j}\right\vert \left.  \alpha_{1}^{i}%
\right\rangle \left\langle \alpha_{2}^{i}\right\vert \left.  \beta_{2}%
^{j}\right\rangle ~~.\label{mh1}%
\end{equation}
As before the notation $\left.  \left\vert \alpha_{1}^{i}\right.
\right\rangle ,\left.  \left\vert \beta_{1}^{j}\right.  \right\rangle $ etc.
refers to the spinor part only of the wavefunction of the particle. Here
$\left\vert V_{12}\right\vert =\frac{\Delta_{01}\Delta_{02}}{E_{1}E_{2}%
}U/R_{12}^{3}$ and $R_{12}$ is the distance between the TS. Let us estimate
the transition amplitude $V\left[  \alpha_{1}^{i},\beta_{2}^{j};\beta_{1}%
^{j},\alpha_{2}^{i}\right]  $ in terms of $V_{12}$. By making use of
Eq.(\ref{mh1}) and the completeness condition in the form of $\sum
\limits_{\alpha_{1}^{i_{1}}}\left\vert \left\langle \beta_{1}^{j}\right\vert
\left.  \alpha_{1}^{i}\right\rangle \right\vert ^{2}=1$, we obtain
\begin{align}
\sum\limits_{\alpha_{1}^{i};\alpha_{2}^{i}}\left\vert V\left[  \alpha_{1}%
^{i},\beta_{2}^{j};\beta_{1}^{j},\alpha_{2}^{i}\right]  \right\vert ^{2} &
=\left\vert V_{12}\right\vert ^{2}\sum\limits_{\alpha_{1}^{i}}\left\vert
\left\langle \beta_{1}^{j}\right\vert \left.  \alpha_{1}^{i}\right\rangle
\right\vert ^{2}\sum\limits_{\alpha_{2}^{i}}\left\vert \left\langle \beta
_{2}^{j}\right\vert \left.  \alpha_{2}^{i}\right\rangle \right\vert
^{2}\nonumber\\
&  =\left\vert V_{12}\right\vert ^{2}\label{bur}%
\end{align}
For further estimation, we make the simplest assumption that all the matrix
elements (\ref{mh1}) entering Eq.(\ref{bur}) are equal. The total number of
terms in the left-hand sum is $n^{2}.$ So
\begin{equation}
\left\vert V\left[  \alpha_{1}^{i},\beta_{2}^{j};\beta_{1}^{j},\alpha_{2}%
^{i}\right]  \right\vert =\frac{\left\vert V_{12}\right\vert }{n}%
\label{newampl}%
\end{equation}

Thus, the tunneling amplitude $V\left[  \alpha_{1}^{i},\beta_{1}^{j}%
;\alpha_{2}^{i},\beta_{2}^{j}\right]  $ entering Eq. (\ref{newrc}) is $n$
times smaller as compared with the case of pairs formed by TLS. Therefore, the
factor $n^{2}$ found above for increasing the probability of formation of
resonating pairs is further reduced by a factor $n$. \textit{So, the total
probability to form a resonating pair increases for multilevel systems}
$\xi=n$ \textit{times as compared with TLS's}.

A similar analysis can by carried out when some of the energy levels remain
degenerate or when the transitions between some of them are forbidden. This
case is investigated by using as an example the spectrum of TS with a nuclear
spin $I=1$ when only quadrupolar effects are taken into account (see Fig.
\ref{Fig3}). In order to find the matrix element of the flip-flop transition
let us again make use of Eq. (\ref{mh1}). Here we must differentiate between
three cases.

In the first case, the transition in both TS's of the pair occurs between
levels with $I_{z}=0$. The corresponding states are denoted by their spinor
parts $\alpha^{0}_{1}, \beta^{0}_{1}$ and $\alpha^{0}_{2}, \beta^{0}_{2}$. By
taking into account that the spinor parts remain unchanged it follows that
\begin{equation}
V \left[  \alpha^{0}_{1}, \beta^{0}_{2}; \beta^{0}_{1}, \alpha^{0}_{2}
\right]  = V_{12} \left\langle \alpha^{0}_{1} \right\vert \left.  \beta
^{0}_{1} \right\rangle \left\langle \beta^{0}_{2} \right\vert \left.
\alpha^{0}_{2} \right\rangle = V_{12}~~. \label{00}%
\end{equation}

In the second case, the transition between $I_{z}=0$ states takes place only
in one TS of the pair. In that case one finds
\begin{equation}
V\left[  \alpha_{1}^{0},\beta_{2}^{j};\beta_{1}^{0},\alpha_{2}^{i}\right]
=V_{12}\left\langle \alpha_{1}^{0}\right\vert \left.  \beta_{1}^{0}%
\right\rangle \left\langle \beta_{2}^{j}\right\vert \left.  \alpha_{2}%
^{i}\right\rangle ~~.\label{01}%
\end{equation}
The sum of the square of the matrix elements is
\begin{equation}
\sum\limits_{\alpha_{2}^{i}}\left\vert V\left[  \alpha_{1}^{0},\beta_{2}%
^{j};\beta_{1}^{0},\alpha_{2}^{i}\right]  \right\vert ^{2}=\left\vert
V_{12}\right\vert ^{2}\sum\limits_{\alpha_{2}^{i}}\left\vert \left\langle
\beta_{2}^{j}\right\vert \left.  \alpha_{2}^{i}\right\rangle \right\vert
^{2}=\left\vert V_{12}\right\vert ^{2}~~.\label{01.1}%
\end{equation}
Hereby the condition $\sum\limits_{\alpha_{2}^{i}}\left\vert \left\langle
\beta_{2}^{j}\right\vert \left.  \alpha_{2}^{i}\right\rangle \right\vert
^{2}=1$ has been used. Since the total number of terms in the sum (\ref{01.1})
is two, one can estimate
\begin{equation}
\left\vert V\left[  \alpha_{1}^{0},\beta_{2}^{j};\beta_{1}^{0},\alpha_{2}%
^{i}\right]  \right\vert =\frac{\left\vert V_{12}\right\vert }{\sqrt{2}%
}~~.\label{01.2}%
\end{equation}

In the third case, when the transitions are between levels with nonzero spin
projection in both TS's, one finds
\begin{align}
\sum\limits_{\alpha_{1}^{i};\alpha_{2}^{i}}\left\vert V\left[  \alpha_{1}%
^{i},\beta_{2}^{j};\beta_{1}^{j},\alpha_{2}^{i}\right]  \right\vert ^{2} &
=\left\vert V_{12}\right\vert ^{2}\sum\limits_{\alpha_{1}^{i}}\left\vert
\left\langle \beta_{1}^{j}\right\vert \left.  \alpha_{1}^{i}\right\rangle
\right\vert ^{2}\sum\limits_{\alpha_{2}^{i}}\left\vert \left\langle \beta
_{2}^{j}\right\vert \left.  \alpha_{2}^{i}\right\rangle \right\vert
^{2}\nonumber\\
&  =\left\vert V_{12}\right\vert ^{2}.\label{11}%
\end{align}

The total number of terms in the left-hand sum in Eq. (\ref{11}) is equal to
$2^{2}$ and, therefore,
\begin{equation}
\left\vert V\left[  \alpha_{1}^{i},\beta_{2}^{j};\beta_{1}^{j},\alpha_{2}%
^{i}\right]  \right\vert =\frac{\left\vert V_{12}\right\vert }{2}%
~~.\label{11.1}%
\end{equation}
Using Eqs. (\ref{newrc}),(\ref{00}),(\ref{01.2}),(\ref{11.1}) one estimates
that the probability to form a resonating pair increases by a factor of
$\xi=1+\frac{8}{9\sqrt{2}}\approx1.6$.

Thus, for the case of a nuclear spin $I=1$ the quadrupolar interactions result
in an increase of the probability of finding resonating pairs by a factor
$\xi\approx1.6$ as compared with TLS's. When in an applied magnetic field the
Zeeman splitting is of the order of the quadrupolar interaction, the
probability of forming a resonating pair increases by a factor of $\xi=3$.
Finally, when the Zeeman energy exceeds the quadrupolar splitting, the factor
is $\xi=1$.

The results obtained in this section are based on the fact that a tunneling
system is a multilevel one. In other words, the energy levels of the TS should
be well resolved. Yet, in an ensemble of interacting TS's the energy levels
fluctuate due to spectral diffusion \cite{bh}. When the scale of spectral
diffusion exceeds the quadrupole splitting the transition occurring from the
different levels of the TS's can not be considered as statistically
independent. The scale of spectral diffusion is about $\gamma T,$ $\gamma=PU$
\cite{bh}. So our approach is valid when the quadrupole splitting $b$ is
\begin{equation}
b>\gamma T~~. \label{11.11}%
\end{equation}
Due to a similar reason the Zeeman splitting should obey
\begin{equation}
g\mu B>\gamma T~~. \label{11.12}%
\end{equation}
This condition establishes the minimal value of the $B$ in order that magnetic
field effects show up.

The results obtained in this section are based on the energy level
classification described in Sec. II. This requires that the tunneling
amplitude $\Delta_{0}$ fulfills the relation $\Delta_{0}\ll\Delta,b$. For that
reason, in the above analysis the factor $\frac{\Delta_{01}\Delta_{02}}%
{E_{1}E_{2}}$ entering Eq. (\ref{trtls}) is a small parameter formed by
strongly asymmetric tunneling systems.

\section{Discussion and Conclusions}

We have shown that the strong magnetic field dependence of the electric
susceptibility in ultracold glasses can be understood by taking into account
the interactions of tunneling systems in the presence of nuclear quadrupolar
moment. The essential point is the following: Even by a small applied magnetic
field the number of different energy levels of a tunneling system is
increased. This in turn, modifies the concentration of resonant tunneling
pairs and leads this way to observable effects.

Two kinds of experiments can be related to the present investigation. The
first kind deals with the experimental determination of the real part of the
response of the system to an applied magnetic field. We find for the field
dependence of the static electric susceptibility the right order of magnitude
of the effect. That a too small effect was found in \cite{bodea} may result
from the use of perturbation theory. Another unexplained feature, namely a
plateau in the temperature dependence of the electric susceptibility,
sometimes called dielectric saturation is most probably due to a dependence of
the tunneling matrix element $\Delta_{0}$ on the spinors of the right and left well.

The second group of experiments is related to measurements of the imaginary
part of the response. In this case information on the relaxation rate of
elementary excitations can be obtained. For example, measuring the echo
amplitude allows to determine the behavior of the TS coherence time $\tau_{2}$
as function of field. The echo amplitude has been found to show a
non-monotonic dependence on the magnetic field \cite{4,5,5a,5b,5c,5d,5e}. It
is defined by the transverse relaxation rate $\tau_{2}^{-1}$. In Sec. 5 we
have introduced the parameter $\xi$ depending on the magnetic field which
controls the concentration of RP. If the relaxation in the system is due to
the resonant pairs, the transverse relaxation rate is directly proportional to
the concentration of RP. Without the quadrupolar effect, the relaxation rate
$\tau_{2}^{-1}=\gamma^{2}T$ \cite{11}. If the quadrupolar effects are taken
into account, the concentration of RP increases by the factor $\xi.$
Therefore, the relaxation rate becomes
\begin{equation}
\tau_{2}^{-1}\approx\xi\gamma^{2}T~~. \label{tau2}%
\end{equation}
The non-monotonic behavior of the parameter $\xi$ on the magnetic field
correlates with the behavior of the echo amplitude found in the experiments
mentioned above. Note that in high magnetic fields $\xi$ reduces to $\xi=1,$
because in that limit the energy levels are equally spaced as in Fig
(\ref{Fig2}).

The standard scheme of interpreting the echo experiments is based on the
assumption that TS's are two-level ones. In that case they can be described in
terms of $\tau_{2}^{-1}$. This is not the case when the quadrupolar effects
are taken into account. Therefore it is no surprise that the calculated
$\tau_{2}^{-1}$ can only qualitatively but not quantitatively describe the
echo experiments. Instead the approach in \cite{9, wurger2004, parshin2004}
based on a multilevel description of TS's instead of the standard Bloch
equations seems very reasonable.

For this reason, another kind of low-temperature experiments should be made to
clarify the role if the $R^{-3}$ interaction between the tunneling system. For
example, measurements of the magnetic field dependence of dielectric loss or
internal friction would allow to extract information on the lifetime $\tau
_{1}$ of the excitations. We want to show that in this case the factor $\xi$
should show up in an even more important way.

The spectral diffusion reaches its maximal value $\gamma T$ at time $\tau
_{1}.$ Therefore, the rate of the spectral diffusion is $\gamma T/\tau_{1}.$
On the other hand the spectral diffusion rate relates to $\tau_{2}^{-2}$~
\cite{11,bh}. Then
\begin{equation}
\tau_{1}^{-1}=\xi^{2}T\gamma^{3} ~~.\label{tau1}%
\end{equation}
Thus the relaxation rate $\tau_{1}^{-1}$ shows an even stronger quadrupole and
magnetic field dependence than $\tau_{2}^{-1}$ and investigation of dielectric
loss or internal friction in the presence of the magnetic field opens an
attractive opportunity to investigate the role of the $R^{-3}$ interaction.

The above analysis is based on the assumption that the EFG has approximately
axial symmetry, i.e., that the parameter $\varkappa$ is small. Generally this
is not the case. Nevertheless, the magnetic field induced relaxation mechanism
has a quite universal character. Indeed, let the nuclear spin be half-integer.
It follows from Kramers' theorem that in zero magnetic field the energy
spectrum is degenerate. So, after application of the magnetic field the total
number of levels of the multilevel TS increases and an effect takes place.

The nature of the tunneling systems in amorphous solids remains puzzling
despite of the large theoretical efforts probing various models. The main
problem of the theory is the lack of experiments which test particular models
versus the original phenomenological model \cite{6}, which simply employs the
distribution (\ref{distrib}). We expect that the magnetic field experiments
allow to reveal which and how many atoms participate in the tunneling. This
should shed new light on the microscopic nature of the tunneling systems in glasses.

\begin{acknowledgments}
The authors would like to thank S. Hunklinger, Yu. Kagan and L.A. Maksimov for
valuable discussions. We thank Mrs. Regine Schuppe for carefully preparing the
manuscript. The work of I. Ya. Polishchuk has been supported through the
Russian Fund for Basic Research (grant 04-02-17469). The work of A. L. Burin,
Y. Sereda and D. Balamurugan is supported by TAMS GL fund (account no. 211043)
through the Tulane University, College of Liberal Arts and Science
\end{acknowledgments}


\begin{thebibliography}{99}                                                                                               %


\bibitem {1}P. Strehlow, C. Enss, and S. Hunklinger, Phys. Rev. Lett.
\textbf{80}, 5361 (1998)

\bibitem {1a}F. Penning, M. Maior, P. Strehlow, S Weigers, H. van Kempen, and
J. Maan, Physica (Amsterdam) \textbf{211B}, 363 (1995)

\bibitem {2}P. Strehlow, M. Wohlfahrt, A.~G.~M. Jansen, R. Haueisen, G. Weiss,
C. Enss and S. Hunklinger, Phys. Rev. Lett. \textbf{84}, 1938 (2000)

\bibitem {3}P. Strehlow, M. Wohlfahrt, C. Enss, and S. Hunklinger, Europhys.
Lett. \textbf{56}, 690 (2001)

\bibitem {4}R. Haueisen, G. Weiss, Phys. \textbf{B 316\&317}, 555 (2002)

\bibitem {5}J.~Le Cochec, F. Ladieu, and P. Pari, Phys. Rev. \textbf{B 66},
064203 (2002)

\bibitem {5a}S. Ludwig, C. Enss, S. Hunklinger, P. Strehlow, Phys. Rev. Lett.
\textbf{88}, 075501 (2002)

\bibitem {5b}C. Enss, S. Ludwig, Phys. Rev. Lett. \textbf{88}, 075501 (2002)

\bibitem {5c}C. Enss, Physica \textbf{B, 316-317}, 12 (2002)\textbf{\ }

\bibitem {5d}S. Ludwig, P. Nagel, S. Hunklinger, and C. Enss, J. Low. Temp.
Phys. \textbf{131}, 89 (2003)

\bibitem {5e}P. Nagel, A. Fleischmann, S. Hunklinger, and C. Enss, Phys. Rev.
Lett. \textbf{92}, 245511 (2004)

\bibitem {dipole-echo}G. Baier, M.~v. Schickfus, Phys.Rev. \textbf{38}, 9952 (1988)

\bibitem {6}P.~W. Anderson, B.~I. Halperin, C.~M. Varma, Philos. Mag.
\textbf{25}, 1 (1972); W.~A. Phillips, J. Low Temp. Phys.\textbf{7}, 351 (1972)

\bibitem {7}S. Hunklinger, A.~K. Raychaudchari, Progr. Low Temp. Phys.
\textbf{9}, 267 (1986)

\bibitem {8}W.~A. Phillips, Rep. Prog. Phys. \textbf{50}, 1657 (1987)

\bibitem {10a}A.L. Burin, Yu. Kagan, JETP \textbf{80}, 761 (1995)

\bibitem {10}\textit{Tunneling Systems in Amorphous and Crystalline Solids,
edited by P.Esquinazi (Springer, Berlin 1998)}

\bibitem {11}A.~L. Burin, Yu. Kagan, L.~A. Maksimov, I.~Ya. Polishchuk, Phys.
Rev. Lett. \textbf{80}, 2945 (1998)

\bibitem {12}A.~L. Burin, Yu. Kagan, I.~Ya. Polishchuk, Phys. Rev. Lett.
\textbf{86}, 5616 (2001)

\bibitem {13}A.~L. Burin, Yu. Kagan, L.~A. Maksimov, I.~Ya. Polishchuk, Phys.
Rev. B \textbf{69} 220201(R) (2004)

\bibitem {14}A.~ L. Burin, I.~Ya. Polishchuk, J. Low. Temp. Phys, 137 (3-4):
189, (2004); arXiv:cond-mat/0407377 v1 15 JUl 2004

\bibitem {8a}S. Kettemann, P. Fulde, and P. Strehlow, Phys. Rev. Lett.
\textbf{83}, 4325 (1999)

\bibitem {8b}A.~W\"{u}rger, Phys. Rev. Lett. \textbf{88}, 075502 (1999)

\bibitem {9}A.~W\"{u}rger, A. Fleischmann, C. Enss, Phys. Rev. Lett.
\textbf{89}, 237601 (2002)

\bibitem {tup}Tupitsyn I.~S., Prokofev N.~V, Stamp P.C.E. Int. J. Mod. Phys. B
11 2901 (1997)

\bibitem {15}A. Abraham, The Principles of Nuclear Magnetism, Oxford 1961

\bibitem {bh}J.~L. Black, B.~I. Halperin, Phys. Rev. B \textbf{16}, 2819 (1968)

\bibitem {jakle}J. J\"{a}ckle, L. Piche, W. Arnold, S. Hunklinger, J.
Non.-Cryst. Sol., \textbf{20}, 365 (1976).

\bibitem {hunk}S. Hunklinger, A.~K. Raychaudchary, Progr. Low Temp. Phys.
\textbf{9}, 267 (1986).

\bibitem {footnote1}To obtain the quadrupole Hamiltonian for the right well
one should put $\theta=0$ in Eq. (\ref{f30.0}).

\bibitem {opposite}The oposite case will be examined in a separate paper.

\bibitem {bodea}D. Bodea, A. W\"{u}rger, J. Low. Temp. Phys, \textbf{136, }39, (2004)

\bibitem {bfp}A.~ L. Burin, I.~Ya. Polishchuk, P. Fulde (to appear)

\bibitem {wurger2004}A.~W\"{u}rger, J. Low. Temp. Phys, 137 (3-4): 143, (2004)

\bibitem {parshin2004}D.A. Parshin, J. Low. Temp. Phys, 137 (3-4): 189, (2004)
\end{thebibliography}
\end{document}